\documentclass[lettersize,journal]{IEEEtran}

\IEEEoverridecommandlockouts
\usepackage{cite}
\usepackage{float}
\usepackage{amsmath,amssymb,amsfonts}
\usepackage{algorithmic}
\usepackage{graphicx}
\usepackage{textcomp}
\usepackage{xcolor}
\usepackage{setspace}
\usepackage{tcolorbox}
\definecolor{appdarkred}{HTML}{E51400} 
\usepackage{booktabs}
\def\BibTeX{{\rm B\kern-.05em{\sc i\kern-.025em b}\kern-.08em
    T\kern-.1667em\lower.7ex\hbox{E}\kern-.125emX}}
\newtcbox{\highlightappdarkred}[1][appdarkred]{
  on line,
  arc=0pt,
  outer arc=0pt,
  colback=#1,
  colframe=black,
  boxrule=0.5mm,
  boxsep=0pt,
  left=1mm,
  right=1mm,
  top=.25mm,
  bottom=.25mm
}

\usepackage{comment}
\usepackage{setspace}
\usepackage{textcomp}
\usepackage{xcolor}
\usepackage{pifont}

\usepackage{fancyhdr}
\usepackage[norelsize, linesnumbered, ruled, lined, boxed, commentsnumbered, noend]{algorithm2e}

\usepackage{hyperref}
\usepackage{cleveref}

\begin{document}
\pagenumbering{arabic}
\title{Scientific Workflow Scheduling in Cloud Considering Cold Start and Variable Pricing Model}

\author{\IEEEauthorblockN{Suvarthi Sarkar, Sparsh Mittal, Shivam Garg, Aryabartta Sahu \textit{IEEE Senior Member}}\\
\IEEEauthorblockA{\textit{Dept. of CSE,  IIT Guwahati, Assam, India.} Emails:\{s.sarkar, m.sparsh, shivam.g, asahu\}@iitg.ac.in }
}

\maketitle

\thispagestyle{plain}
\pagestyle{plain}

\begin{abstract}
Cloud computing has become a pivotal platform for executing scientific workflows due to its scalable and cost-effective infrastructure. Scientific Cloud Service Providers (SCSPs) act as intermediaries that rent virtual machines (VMs)  from Infrastructure-as-a-Service (IaaS) providers to meet users' workflow execution demands. The SCSP earns profit from the execution of scientific workflows if it completes the execution of the workflow before the specified deadline of the workflow. This paper addresses two key challenges that impact the profitability of SCSPs: the cold start problem and the efficient management of diverse VM pricing models, namely reserved, on-demand, and spot instances. 

We propose a hybrid scheduling framework that integrates initial planning based on historical data with real-time adaptations informed by actual workload variations. In the initial phase, VMs are provisioned using reserved pricing based on predicted workloads and spot instances. During execution, the system dynamically adjusts by provisioning additional VMs through on-demand or spot instances to accommodate unexpected bursts in task arrivals. Our framework also incorporates a dependency-aware task scheduling strategy that accounts for cold start delays and spot pricing volatility.
Experimental results on real-world benchmark datasets demonstrate that our approach outperforms state-of-the-art methods, achieving up to 20\% improvement over cold-start-focused techniques and 15\% over pricing-model-based VM provisioning strategies. 

\end{abstract}


\section{Introduction}

Cloud computing has become essential for executing diverse workloads and workflows due to its inherent flexibility, scalability, and cost-effectiveness. Users typically submit workflows to cloud environments, which then provision computational resources necessary for their execution \cite{yu2017survey, rodriguez2017taxonomy}. This research specifically focuses on scientific workflows, referring to cloud infrastructure providers as Scientific Cloud Service Providers (SCSPs).

The SCSP functions primarily as a broker, receiving scientific workflows from users and renting virtual machines (VMs) from Infrastructure-as-a-Service (IaaS) providers. This approach allows SCSPs to avoid the capital expenditures (CapEx) associated with owning physical infrastructure. The primary objective of the SCSP is profitability, achieved by charging users subscription or usage fees and strategically utilising these funds to rent resources from IaaS providers. Notable examples of IaaS providers include Amazon Web Services (AWS EC2), Microsoft Azure, and Google Cloud Platform (GCP). Examples of SCSPs are Jetstream, Chameleon Cloud, Open Science Grid, CloudLab, and SciServer.

In addressing this objective, two critical challenges are highlighted:
First, the cold start problem significantly complicates workflow execution. Cold start refers to delays encountered during initial setup before task execution begins, typically accounting for about 20\% of total execution time \cite{cold_start}. Such delays can degrade performance, lead to inefficient resource use, and increase the likelihood of deadline violations. This issue is especially critical in online workflow scheduling, where task arrivals are dynamic and unpredictable, necessitating intelligent scheduling strategies that balance cost and execution efficiency. Moreover, this problem is especially important to address because studies have shown that a small fraction of functions, approximately 20\%, are invoked around 99\% of the time \cite{cold_start}. Thus, by strategically leveraging caching mechanisms and optimizing cold start handling, substantial improvements in execution time and resource utilization can be achieved.

The second challenge involves strategically renting machines under various pricing models offered by IaaS providers \cite{zhao14}. These providers typically offer multiple VM types with differing memory and computational capacities, rentable through reserved, on-demand, or spot pricing models \cite{kumar18}. Reserved instances are booked in advance based on historical usage data, providing predictable costs but risking either underutilization or insufficient resources during unexpected workloads. Conversely, on-demand instances, though more expensive, accommodate bursts of real-time demand. On the other hand, spot instances are offered at significantly discounted prices and are made available by IaaS in real-time based on current supply and demand. However, their availability is not guaranteed, and they may not always be accessible when needed. Spot instances offer real-time, discounted pricing but come with reliability risks. The SCSP needs to bid for it, and spot VM availability can be revoked anytime if the market price exceeds the SCSP's bid. This uncertainty may interrupt tasks, requiring reallocation and re-execution. Despite this, their low cost makes them attractive for non-critical or fault-tolerant tasks \cite{SINGH2022}.

The main challenge is finding the right balance between different VM pricing options. Reserving too many VMs can waste money if they are not fully used, but they help reduce cold start delays because caching works better on reserved machines. If fewer VMs are reserved, the system may need to use expensive on-demand VMs. Spot VMs are cheaper but come with a risk. If the price goes above the bid, the VM is taken away. Bidding higher avoids this but increases cost; bidding lower saves money but risks losing the VM. Our work focuses on solving this problem to help the SCSP make more profit by: (a) managing many workflows from users, (b) using spot price changes wisely, (c) reducing cold start costs, and (d) planning based on past data and changing prices.

To address these challenges, we propose a hybrid two-phase scheduling strategy. In the first phase, we develop a preliminary schedule based on historical workload data (referred to as the first phase or predicted case). Subsequently, in the second phase, adjustments are made dynamically in real-time (referred to as the second phase or actual/ predicted case) to accommodate deviations from the predicted workload. Critical challenges during real-time scheduling include managing cold start delays, dynamically adjusting reserved bookings, and effectively utilizing on-demand and spot instances despite the inherent risk of spot instance revocations. These challenges are thoroughly addressed in our real-time scheduling framework.

The main contributions of this work are as follows:

\begin{itemize} \item We propose an integrated scheduling framework that jointly addresses cold start delays, task dependencies, and VM pricing strategies—unlike prior approaches that treat these concerns independently. To the best of our knowledge, this is the first work to incorporate all three major VM pricing models (reserved, on-demand, and spot) in a unified scheduling strategy.

\item We introduce a hybrid two-phase scheduling strategy: an offline phase that plans reserved VM usage based on predicted workloads and spot availability, and a real-time phase that dynamically provisions on-demand and spot instances in response to actual workflow demands.

\item We incorporate deadline- and dependency-aware scheduling by proportionally distributing workflow deadlines among tasks. This approach minimizes cold start overheads and reduces deadline violations, directly contributing to higher profit margins.

\item We develop a cost-aware VM pool management strategy that includes a priority-based VM selection mechanism and a reward-guided spot bidding policy, both of which optimize task placement while mitigating the risk of spot instance revocation.

\item Experimental evaluations using real-world scientific workflows and pricing traces demonstrate consistent performance improvements over state-of-the-art baselines. Our framework remains robust under varying workload intensities, spot availability patterns, and pricing dynamics, achieving up to 80.394\% profit even with up to 40\% prediction error. \end{itemize}

\section{Literature Review}
\nocite{6,7}
\begin{table}[tb!]
    \centering
    \caption{Literature Review Summary. CS stands for cold start, RC stands for reserved cost, DEP stands for dependency, and SC stands for Spot cost. }
    \begin{tabular}{|c|c|c|c|c|c|c|} \hline
         Paper & CS & RC & SC & DEP & Online  \\ \hline
         Fuerst et al. \cite{10.1145/3445814.3446757} & \ding{51} & \ding{55} & \ding{55} & \ding{55} & \ding{51} \\
         Suo et al. \cite{9556063} & \ding{51} & \ding{55} & \ding{55} & \ding{51} & \ding{51} \\
         Nezamdoust et al. \cite{10.1007/s11227-022-04970-x} & \ding{55} & \ding{55} & \ding{51} & \ding{55} & \ding{55} \\
         Taghavi et al. \cite{10.1007/s10723-023-09676-9} & \ding{55} & \ding{55} & \ding{51} & \ding{51} & \ding{51} \\         
         Proposed work & \ding{51} & \ding{51} & \ding{51} & \ding{51} & \ding{51} \\ \hline
    \end{tabular}
    
    \label{tab:lit rev}
\end{table}
Maximizing profit for cloud service providers requires balancing performance and cost efficiency. Two major factors are cold start mitigation and the strategic use of VM pricing models. While previous work has explored these challenges separately, many overlook key aspects such as workflow dependencies in cold start scenarios or the joint use of reserved, on-demand, and spot instances. This section reviews literature on cold start reduction and cloud pricing strategies.
\subsection{Cold Start Literature}

\nocite{3,4,5}
Researchers have invested significant effort into addressing the cold start issue in Function-as-a-Service (FaaS) platforms in recent years. Since a cold start involves creating a container, fetching, and installing necessary libraries and dependencies before executing the function itself, the resulting latency is often several times—or even tens of times—higher than that of a warm start, as discussed by Li et al. \cite{10.1145/3508360} and Pu et al. \cite{227653}. 
Researchers have explored various techniques to mitigate cold starts in cloud computing, including keep-alive policies, pre-warming strategies, and dependency-aware scheduling. While early efforts like those by Suo et al. \cite{9556063} and Shahrad et al. \cite{cold_start} focused on minimizing the number of cold starts, they often overlooked total cold start latency and workflow dependencies.

Cache-based solutions have emerged as a promising direction. Fuerst et al. \cite{10.1145/3445814.3446757} introduced a dual-caching eviction policy to reduce cold start delays, while Romero et al. \cite{10.1145/3472883.3486974} proposed a distributed cache with transparent auto-scaling. However, most cache strategies do not fully account for workflow task dependencies, limiting their effectiveness. Cai et al. \cite{10494685} aimed to minimize total cold start time but did not leverage dependency information—an important factor for further optimization.

\nocite{1,2}
\subsection{Literature on Pricing Strategies }

Several studies have investigated the use of spot instances to reduce operational costs, given their significantly lower prices compared to on-demand resources. For example, Nezamdoust et al. \cite{10.1007/s11227-022-04970-x} proposed a deep learning-based approach for forecasting spot instance prices, enabling more informed bidding strategies. However, their work focuses solely on price prediction and does not address the challenge of balancing reserved, on-demand, and spot instances for holistic cost optimization. Similarly, Malpuri et al. \cite{articlespotfleet} briefly highlight the advantages of combining spot, on-demand, and reserved instances but fall short of providing a concrete strategy for managing trade-offs among these options or offering insights into optimal bidding for spot instances. Taghavi et al. \cite{10.1007/s10723-023-09676-9} proposed a cost-efficient broker model for Workflow-as-a-Service (WaaS), utilizing on-demand and spot instances to reduce execution costs. Chen et al. \cite{8066736} explored spot block and on-demand instances for tasks that allow preemption. While these approaches effectively leverage dynamic pricing, they overlook the potential benefits of incorporating reserved instances, which can yield additional savings for workloads with partially predictable behaviour.

While existing literature has extensively studied cold start mitigation and VM pricing models individually, a unified scheduling framework that jointly considers cold start delays, task dependencies, and all three major VM pricing models (reserved, on-demand, and spot) remains largely unexplored. In this work, we address this gap in the existing literature. A summary of the reviewed literature is shown in \autoref{tab:lit rev}.

\section{System Model and Problem
Statement}
\label{sec:prob}
\subsection{System Model}
The Scientific Cloud Service Provider (SCSP) receives a continuous stream of scientific workflows, which it processes in batches. These workflows are temporarily collected in a pool and scheduled periodically. This batch-wise scheduling strategy is effective, particularly because workflow execution times are typically longer than the time required for scheduling, and global information aids in more efficient scheduling decisions.

The SCSP does not own the computational resources to execute the workflows. Instead, it rents Virtual Machines (VMs) from an Infrastructure-as-a-Service (IaaS) provider. The IaaS provider maintains a diverse pool of VM types and offers them to the SCSP on a pay-per-use basis.

The SCSP earns a reward only when a workflow is successfully completed within its deadline. Therefore, it must manage its VM rentals efficiently by strategically selecting VM types at specific prices, minimizing idle costs, and leveraging cold start delays to maximize the overall profit.

\begin{figure}[tb!]
    \centering
    \includegraphics[width=1\linewidth]{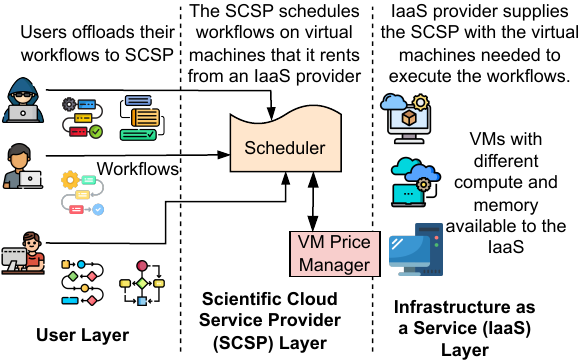}
    \caption{Considered System Architecture}
    \label{fig:architecture}
\end{figure}

 A VM Price Manager component continuously updates the current spot prices for different VM types from the IaaS provider, enabling the SCSP to make cost-effective and informed decisions when renting VMs. The overall system architecture is shown in \autoref{fig:architecture}.

\subsection{Workflow Model}

Let $\mathcal{W}$ denote the set of $K$ workflows received by the SCSP in a given batch, such that $\mathcal{W} = {W_1, W_2, \cdots, W_k, \cdots, W_K}$. Each workflow $W^k$ is represented as a Directed Acyclic Graph (DAG) $W^k = (V^k, E^k)$, where $V^k$ is the set of tasks and $E^k$ represents the precedence constraints between them.

Specifically, an edge $\{v_{i_1}^k, v_{i_2}^k\} \in E^k$ implies that task $v_{i_1}^k$ must complete before task $v_{i_2}^k$ can begin execution. The set $V^k$ contains $n^k$ tasks, and each workflow $W^k$ is associated with an arrival time $a^k$, a deadline $d^k$, and a reward $r^k$ earned upon successful completion within its deadline. The reward model is adopted as $r^k =  l_i^k * \{ \frac{l_i^k}{\text{Length of Critical Path of } W^k} \}^2 $, following the formulation given in \cite{SwainGS20}. 

Each task $v_i^k \in V^k$ is defined as a 3-tuple $(l_i^k, m_i^k, c_i^k)$, where, $l_i^k$ denotes the task length measured in millions of instructions (MI), $m_i^k$ represents the memory requirement of the task, and $c_i^k$ is the cold start length, i.e., the length of instructions needed to load the execution environment.
Tasks can be scheduled for execution based on these parameters, considering both the precedence constraints and resource requirements.

\subsection{Cold Start Model}

Virtual Machines (VMs) provide isolated environments comprising an operating system and a set of dependencies required to run specific applications. These dependencies are often task-specific and can only be reused if the same task type appears again, either within the same workflow or across different workflows.

A cold start in a VM refers to the delay incurred when the VM is initialized for the first time. This delay includes booting the operating system, allocating resources, and loading required dependencies and configurations. The cold start delay depends on both the host VM and the task. Leveraging the cold start model implies minimizing this delay by reusing the already loaded environment for tasks of the same type.

 The total execution time of a task $v_i^k$ on a VM $j$, denoted by $t^k_{ij}$, depends on two components: the task's length ($l_i^k$) and the cold start time in terms of length ($c_i^k$). The execution time is given by:

\begin{equation}
    t^k_{ij} = \frac{l_i^k}{CP_j} + y_{ij}^k \cdot \frac{c_i^k}{CP_j}
\end{equation}

Here, $\frac{l_i^k}{CP_j}$ represents the time required to handle the computation requirement of the task, $CP_j$ is the computational power of $VM_j$, $c_i^k$ is the cold start time needed to initialize the environment for task $v_i^k$, $y_{ij}^k$ is a binary indicator that captures whether a cold start is needed. The variable $y_{ij}^k = 1$, if the previously executed task on $VM_j$ is of a different type (cold start occurs), $y_{ij}^k = 0$ if the previous and current tasks are of the same type (cold start is avoided).
By reusing environments for tasks of the same type, the system can significantly reduce execution latency and improve resource efficiency.

\subsection{Cost Model}

We consider three VM renting cost models—Reserved, On-Demand, and Spot—used by the Scientific Cloud Service Provider (SCSP) to rent VMs from an Infrastructure-as-a-Service (IaaS) provider. These VMs are utilized for executing user-submitted workflows. Although the SCSP can rent unlimited VM instances, only a finite set of VM configurations is available. Each VM instance $VM_j$ has a configuration described by the tuple $(CP_j, mem_j, rt_j, rc_j)$, where, $CP_j$ represents computational power in terms of MIPS, $mem_j$ represents memory capacity, $rt_j$ represents the renting duration, $rc_j$ represents renting cost of the VM per unit time (it is $R$ for reserved type, $D$ for demand cost, bidding price for spot case). The renting type can be categorized into one of the following three types.

\subsubsection{Reserved Pricing}
Reserved instances offer lower costs by allowing users to pre-book VMs for extended durations. They are ideal for predictable, long-running workloads due to their lower per-unit cost than On-Demand instances. However, they require upfront payment, irrespective of actual VM utilization. The total reserved cost of SCSP is denoted by $C_{\text{res}}$:
\begin{equation}
C_{\text{res}} = \sum_{j=1}^{M} \alpha_j \cdot rc_j (=RP) \cdot rt_j
\end{equation}
where $\alpha_j$ is a binary indicator variable (1 if $VM_j$ is reserved, 0 otherwise), $M$ is the total number of VM instances.

\subsubsection{On-Demand Pricing}
On-Demand instances offer flexibility by allowing users to rent VMs instantly without any long-term commitment. They are well-suited for unpredictable or bursty workloads but come at a higher per-unit cost than reserved instances. This model is beneficial when immediate computational resources are required in response to sudden or emergency demands. The total on-demand cost ($C_{\text{dem}}$) is:
\begin{equation}
C_{\text{dem}} = \sum_{j=1}^{M} \beta_j \cdot rc_j (=DP) \cdot rt_j
\end{equation}
where $\beta_j$ is a binary indicator variable (1 if $VM_j$ is rented on demand, 0 otherwise).

\subsubsection{Spot Pricing}
Spot instances offer significant cost savings by allowing SCSPs to bid for spare VM capacity at dynamically changing prices. We denote the spot bidding price for $VM_j$ as $bid\_price_j$. However, these instances may be revoked anytime if the current spot price (denoted by $SP$) exceeds the SCSP's bid price. If an instance is terminated, the last computed state is stored, allowing the task to resume execution when another VM becomes available.

The cost for spot instances ($C_{\text{spot}}$) is calculated as:
\begin{equation}
C_{\text{spot}} = \sum_{j=1}^{M} \gamma_j \cdot rc_j (=bid\_price_j) \cdot rt_j
\end{equation}
where $\gamma_j$ is a binary indicator (1 if $VM_j$ is a spot machine, 0 otherwise).

When comparing the three cost models for VM with the same memory capacity, compute capacity and rental duration, the cost per unit time is highest for demand instances, followed by reserved cases, and lowest for spot instances.

Hence, the total VM renting cost ($C$) is:
\begin{equation}
C = C_{\text{res}} + C_{\text{dem}} + C_{\text{spot}}
\end{equation}

For simplicity, we consider that if a specific VM is hired multiple times, we consider a different VM index for each hiring. Also, after the rental period is exhausted, the SCSP can not leverage the advantage of a cold start for that VM.

\subsection{Problem Statement}

We define the start time of the $i^{th}$ task of the $k^{th}$ workflow as $st_i^k$ and the finish time as $ft_i^k$. So, it can be written as $ft_i^k =  st_i^k + t_i^k, \: \forall \: v_{i}^{k} \in V^{k}$. We define an indicator variable $z^k$ that denotes whether the $k^{th}$ workflow met the deadline $d^{k}$ as follows
\[
z^k = 
\begin{cases} 
1 & \text{if } \max\limits_{\forall i} \;( ft_i^k )\: \leq \: d^k \\
0 & \text{otherwise}
\end{cases}
\]

Our problem can be modelled as an optimization problem with the following objective function.
\begin{equation}
    Profit = max (\sum_{k=1}^K r^k \cdot z^k  \: - \:  C )
\end{equation}

Here, the total profit depends on two terms. The first term is the total reward accumulated by completing the workflows within the deadlines, and the second term is the total cost incurred for renting all the machines. 

This is subject to the following constraints:
\begin{equation}
ft_{i_1}^k \leq st_{i_2}^k \; \; \forall \;  \{v_{i_1}^k , v_{i_2}^k\} \in E^k ,\: \forall k
\label{eq: task dependency constraint}
\end{equation}
Constraint (\ref{eq: task dependency constraint}) is the task dependency constraint ensuring that each task starts executing after all its predecessors have finished execution.

We define a binary indicator variable $x_{ij}^{k}$, which takes the value 1 if task $v_i^k$ is assigned to VM $j$, and 0 otherwise.
\begin{equation}
    \sum_{j}^{M} x_{ij}^{k} \leq 1 , \forall i ,\: \forall k
    \label{eq: max one VMconstraint}
\end{equation}
Constraint (\ref{eq: max one VMconstraint}) ensures that each task is executed on at most one VM.
\begin{equation}
    \sum_{j}^{M} x_{ij}^{k}\cdot mem_{j} \geq m_{i}^{k}, \; \forall i \; \forall k
    \label{eq: min memory constraint}
\end{equation}
Constraint (\ref{eq: min memory constraint}) ensures that the memory of the VM on which a task is scheduled is greater than or equal to the amount of memory required by the task.

\begin{align}
&\text{If }  st^{k_1}_{i_3} < st^{k_2}_{i_4},\: x^{k_1}_{i_3j} = x^{k_2}_{i_4j} =1 \nonumber \\
&\text{then } x^{k_1}_{i_3j} \cdot ft^{k_1}_{i_3} \leq x^{k_2}_{i_4j} \cdot st^{k_2}_{i_4}
\quad \forall i,\; \forall j,\; \forall k
\label{eq: VMdoes atmost one task}
\end{align}

Constraint (\ref{eq: VMdoes atmost one task}) ensures that each VM executes at most one task at any given time. Specifically, if two tasks $v^{k_1}_{i_3}$ and $v^{k_2}_{i_4}$ are scheduled consecutively on the same VM $j$, the finish time of $v^{k_1}_{i_3}$ must not overlap with the start time of $v^{k_2}_{i_4}$.

\begin{equation}
    \max_{k}^{K}(\max_{i}^{n^k} \;  (x_{ij}^k \cdot ft_i^k)) - \min_{k}^{K}(\min_{i}^{n^k} \;  (x_{ij}^k \cdot st_i^k)) \leq rt_j,  \; \forall j
    \label{eq: total available time constraint}
\end{equation}
Constraint (\ref{eq: total available time constraint}) states that the total time each VM runs is less than or equal to the total time for which the VM was rented.
\begin{equation}
    \alpha_{j} + \beta_{j} + \gamma_{j} = 1 
    \label{eq: VMtype constraint}
\end{equation}
Constraint (\ref{eq: VMtype constraint}) states that each machine can either be procured through one of the three policies, reserved, on-demand or spot. We summarise all the symbols used in \autoref{tab:notation} for ease of reading.

\begin{table}[tb!]

\centering
\footnotesize
\caption{Notations used}
\begin{tabular}{|p{1.2cm}|p{6.8cm}|}
\hline
\textbf{Symbol} & \textbf{Meaning} \\ \hline
$W^k$ & $k^{th}$ workflow \\ 
$K$ & The total number of workflows\\
$n^{k}$ & Total number of tasks in $k^{th}$ workflow \\
$E^k$ & Dependency graph of $k^{th}$ workflow where nodes are tasks \\ 
$V^k$ & Set of tasks in $k^{th}$ workflow \\ 
$d^k$ & Deadline of $k^{th}$ workflow \\ 
$r^k$ & Reward on completing $k^{th}$ workflow before deadline $d^{k}$\\ 
$z^k$ & 1 if $k^{th}$ workflow finishes before deadline $d^{k}$ \\ 
$v_i^k$ & $i^{th}$ task in $k^{th}$ workflow \\ 
$l_i^k$ & Length of $i^{th}$ task of $k^{th}$ workflow in number of instructions, measured in millions of instructions\\ 
$m_i^k$ & Memory requirement of $i^{th}$ task of $k^{th}$ workflow \\ 
$c_i^k$ & Cold start time of $i^{th}$ task of $k^{th}$ workflow \\ 
$st_i^k$ & Start time of $i^{th}$ task of $k^{th}$ workflow \\
$ft_i^k$ & Finish time of $i^{th}$ task of $k^{th}$ workflow \\
$rd^k_i$ & Relative deadline of $i^{th}$ task of $k^{th}$ workflow \\
$rcp^k_i$ & Relative compute power of $i^{th}$ task of $k^{th}$ workflow \\
$t_i^k$ & Total execution time of $i^{th}$ task of $k^{th}$ workflow including cold start time \\ 
$y_{ij}^k$ & 1 if there is a cold start for $i^{th}$ task of $k^{th}$ workflow on $vm_j$ \\
$x_{ij}^{k}$ & 1 if task $v_i^k$ is assigned to VM $j$, and 0 otherwise \\
$M$ & Total Number of VMs used. \\ 
$vm_j$ & $j^{th}$ VM \\ 
$CP_j$ & Compute power of $j^{th}$ VM in MIPS \\ 
$mem_j$ & Memory available at $j^{th}$ VM \\ 
$rt_j$ & Renting duration of $j^{th}$ VM \\
$\alpha_{j}$ & Indicator variable, indicating whether the $j^{th}$ machine is reserved. \\
$\beta_{j}$ & Indicator variable, indicating whether the $j^{th}$ machine is procured on-demand.\\
$\gamma_{j}$ & Indicator variable, indicating whether the $j^{th}$ machine is procured through a spot offering.\\
$C, C_{\text{res}},$ & Total cost, reserved cost\\
$C_{\text{dem}}, C_{\text{spot}}$ & On-demand cost, spot cost\\
\hline
\end{tabular}
\label{tab:notation}
\end{table}

\section{Solutions Approach}
\begin{figure}
    \centering
    \includegraphics[width=\linewidth]{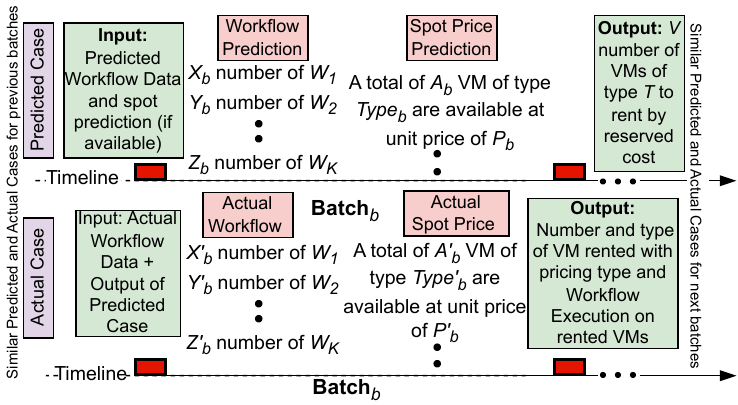}
    \caption{Temporal Pipeline of Proposed Approach. $X_{b}$, $Y_{b}$, $Z_{b}$, $A_{b}$ and $P_{b}$ represent predicted values, which are close but not identical to the actual values $X'_{tb}$, $Y'_{tb}$, $Z'_{b}$, $A'_{b}$, and $P'_{b}$. $b$ represents the corresponding batch number. The approach leveraging predicted workflows and spot instances operates ahead of time based on forecasted data, whereas the approach utilizing actual workflow data executes in real time. The trigger point of our proposed approach is highlighted by \highlightappdarkred{\textcolor{appdarkred}{H}}.}
    \label{fig:timeline}
\end{figure}

\begin{figure}
    \centering
    \includegraphics[width=\linewidth]{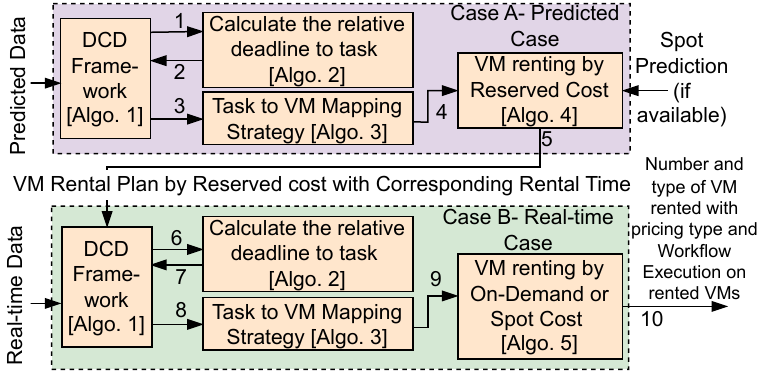}
    \caption{Flowchart of the Proposed Solution Approach for each batch. The numbers represent the ordering of the modules.}
    \label{fig:flowchart}
\end{figure}

We propose a two-phase scheduling approach: Case A (Predicted Case) and Case B (Actual/Real-Time Case). In Case A, the system uses historical data to predict how many workflows of each type will arrive in each batch and plans reserved VM rentals accordingly, including their types and provisioning times. In Case B, the system adjusts this plan in real time by renting additional on-demand or spot VMs as needed. \autoref{fig:timeline} shows the input and output illustration of our proposed approach, while \autoref{fig:flowchart} shows the flow of modules and their interactions, described in later sections.

\subsection{Overall Framework of the Proposed Approach}

The pseudocode of the framework is shown in Algorithm \autoref{alg:framework}. The proposed Deadline, Cold Start, and Dependency-Aware (DCD) Framework continuously schedules tasks across dynamic batch intervals by considering workflow dependencies, task deadlines, cold start overheads, and cost constraints. The framework maintains a priority queue $\mathcal{Q}$ that holds all tasks whose predecessor dependencies have been satisfied. At the beginning of each batch interval, $\mathcal{Q}$ is updated with newly eligible tasks (\texttt{line 5}). For each task $v^k_i$ in $\mathcal{Q}$, a relative deadline $rd^k_i$ is computed (\texttt{line 7}), followed by the estimation of the minimum computational power required to meet the deadline (\texttt{line 8}). A suitable VM is then selected to meet the relative deadline, which optimizes task placement by accounting for cold start penalties, resource configurations, and cost-effectiveness (\texttt{line 10}). Tasks are either scheduled ahead of execution during the prediction phase or hosted in real time as workflows arrive. 
We consider the batch time to be small, in minutes, while the renting time is an hour. 

\begin{algorithm}[tb!]
\label{alg:framework}
\caption{Deadline, Cold Start and Dependency-Aware Approach (DCD) Framework}
\KwIn{\footnotesize $\mathcal{W} \gets$ List of workflows in a batch, VM configurations and costs}
\KwOut{\footnotesize Schedule between tasks of workflow and VM (for predicted case) or workflow execution (real-time case)}

    $\mathcal{Q} \leftarrow \phi$, \texttt{queue contain all tasks with all predecessors tasks executed} \\

    $\mathcal{C} \leftarrow 0$, \texttt{total renting cost}
    
\While{\textnormal{\(true\)}}{
\ForEach{\textnormal{start of batch interval}}{
    $\mathcal{Q} \gets$ all the tasks whose all predecessors are executed

\While{$\mathcal{Q}$ \textnormal{is not empty, do for each task} $v^k_i$}{
    $rd^k_i \gets$ Calculate deadline of the task $v^k_i$ using Algorithm \autoref{alg:deadline}
    ($W^k$, $v^k_i$)

    $rcp^k_i \gets$ Calculate the relative computation required for task $v^k_i$ so that deadline is met
    
    $VM_j \gets$Find machine to allocate using Algorithm \autoref{alg:vmrent} (VM infos, $v^k_i$, $rd^k_i$, $rcp^k_i$, $\mathcal{C}$)

    Task schedule (or host) on $VM_j$

}
\textbf{EndWhile}
}
\textbf{EndFor}
}
\textbf{EndWhile}

\end{algorithm}

\subsection{Calculating Relative Deadlines of Tasks}

\begin{figure}
    \centering
    \includegraphics[scale=1]{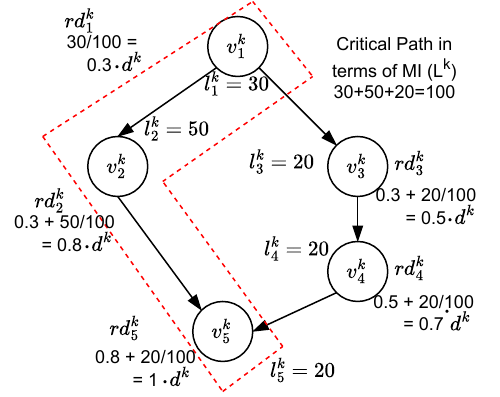}
    \caption{A sample example showing the relative deadline calculation of all tasks in a workflow using \autoref{eqn:reldeadline}. The critical path of workflow $W^k$ is marked inside the dotted line.}
    \label{fig:reldeadline}
\end{figure}

Upon the arrival of a workflow $W^k$, its task dependency ${V^k, E^k}$ is also provided. This enables the preliminary computation of a relative deadline $rd_i^k$ for each task $v_i^k$ within the workflow $W^k$. The pseudocode of relative deadline calculation is outlined in Algorithm \autoref{alg:deadline}. The objective is to complete each task within its assigned relative deadline. We first compute the critical path length of $W^k$ in terms of Millions of Instructions (MIs), denoted as $L^k$ (\texttt{line 2}). The relative deadline for each task is then calculated using \autoref{eqn:reldeadline} (\texttt{line 3}), which proportionally distributes the overall workflow deadline among its constituent tasks.

\begin{align}
L^k &= \sum_{i=1}^{n^k} l^k_i, \quad
rd^k_i &= \max_{\hat{i} \in Pred(i)} rd^k_{\hat{i}}  + \frac{l^k_i}{L^k} \cdot d^k
\label{eqn:reldeadline}
\end{align}
Here, $\max_{\hat{i} \in Pred(i)} rd^k_{\hat{i}}$ represents the maximum relative deadline among the parents of $v^k_i$.
\autoref{fig:reldeadline} illustrates an example of the calculation of relative deadlines for a workflow $W^k$. Here, $L^k$ represents the total critical path length in terms of computational load. For each task $v^k_i$, its share of the total deadline is determined based on the ratio of its load to the overall critical path load. Accordingly, the total workflow deadline is proportionally distributed among the tasks based on their respective load contributions.

\begin{algorithm}[tb!]
\label{alg:deadline}
\caption{Assign Deadlines to Tasks of Workflow}
\KwIn{\footnotesize $W^k \gets$ $k^{th}$ workflow, $v^k_i \gets$ $i^{th}$ task of $k^{th}$ workflow}
\KwOut{\footnotesize Assigned relative deadline $rd_i^k$ for each task $v^k_i$ in all the workflows of batch}

\ForEach{workflow $W^k \in \mathcal{W}$}{

    $L^k = \sum l^k_i \quad \forall \: i \: \text{in} \: \text{critical} \: \text{path} \: \text{of} \: W^k$ 

    $rd^k_i \gets $Calculate relative deadline of the task using \autoref{eqn:reldeadline}

}
Return relative deadline ($rd_i^k$) of task

\end{algorithm}

\subsection{Selection of In-Stock VMs for Executing Tasks}
We first try to schedule the tasks without increasing the pool size of VM, if it is not possible we increase the VM pool by renting new VMs and schedule the task on it. 

For scheduling the task without increasing the pool (selection of in-stocked VMs), we maintain three VM tracking lists: (a) $free\_VMs$ for VMs that are currently rented but idle, (b) $busy\_VMs$ for VMs executing tasks, and (c) $suitable\_VMs$, a temporary subset of $free\_VMs$ that meet the memory and compute requirements of the task being scheduled.

The approach of selecting the in-stocked VM is given in Algorithm \autoref{alg:vmrent}. When selecting a VM for task execution from the existing VM pool, priority is given to avoiding cold starts, as they introduce significant delays. Among VMs that avoid cold starts, we choose those with the smallest memory capacity (\texttt{line 5, 6}). If no such VM is available, we evaluate candidate VM based on a priority score designed to minimize performance penalties (\texttt{line 7, 8}). The priority score is calculated using \autoref{eq:new_priority}.

We consider the priority function based on Zipf's law \cite{zipf1, roy23}, considering the last use timestamp, task popularity, cold start penalty, and memory footprint. The score is computed as:
\begin{equation}
    Priority_j = \psi_1 \cdot LUT_j + \psi_2 \cdot Freq_j * Penalty_j + \psi_3 \cdot mem_j, \: \forall j 
    \label{eq:new_priority}
\end{equation}
Here, $LUT_j$ denotes the last use timestamp, with lower values indicating machines that have not been used recently. $Freq_j$ represents the number of invocations of the most recent task executed on the VM, and $Penalty_j$ is the cold start time associated with that task. $mem_j$ denotes the VM’s memory capacity. Coefficients $\psi_1$, $\psi_2$, and $\psi_3$ balance the relative impact of these components.
We select the VM with the lowest priority score (i.e., highest scheduling preference), schedule the task on it, remove it from $free\_VMs$, and add it to $busy\_VMs$.

If no suitable VMs are available for task assignment, new VMs are rented. In the first phase (predicted case), new VMs are rented using the reserved pricing model as described in Algorithm \autoref{alg:reserved} (\texttt{first part of line 10}). In the second phase (real-time case), VMs are rented by paying on-demand or spot cost following the strategy outlined in Algorithm \autoref{alg:spot1} (\texttt{second part of line 10}). Once new VMs are rented, the pending tasks are scheduled and hosted on these newly provisioned resources.

\begin{algorithm}[tb!]
\label{alg:vmrent}
\caption{Task to VM Mapping Strategy}
\KwIn{\footnotesize VM configurations and costs, $v^k_i \gets$ $i^{th}$ task of $k^{th}$ workflow, relative deadline $rd^k_i$, relative computation power $rcp^k_i$, $\mathcal{C} \leftarrow$ total renting cost}
\KwOut{\footnotesize Selected VMs for task scheduling or hosting}

$free\_VMs \leftarrow$ \texttt{list of VMs rented and free}\\ 
    
$busy\_VMs \leftarrow$ \texttt{list of VMs rented and are in use} \\ 
    
$suitable\_VMs \gets$ select a VM in $free\_VMs$ such that $CP_j \geq rep^k_i$ and $mem_j \geq m^k_i$, ensuring that the selected VM is free and available for the entire execution duration of task $v^k_i$

\If{ \textnormal{one or more VMs in} $suitable\_VMs$}{
    \If{\textnormal{the VM can leverage cold start}}{
        Select the VM with the lowest $CP_j$ and $mem_j$
    }
    \Else{
        Select a VM using the priority function in \autoref{eq:new_priority}
    }
}
\Else{
     \hspace{.2 cm}\texttt{Case A: Predicted case} \newline
Update $free\_VMs$ by renting new VMs using Algorithm \autoref{alg:reserved} ($free\_VMs$, $v^k_i$, $\mathcal{C}$) to schedule $v^k_i$ \newline
    or \newline
    \texttt{Case B: Real-time case} \newline
    Update $free\_VMs$ by renting new VMs using Algorithm \autoref{alg:spot1} ($free\_VMs$, $v^k_i$, $\mathcal{C}$) to host $v^k_i$

    Add the newly rented VM to $busy\_VMs$ by removing it from $free\_VMs$
}
Return the VM to schedule (for predicted case) or host (for real-time case) for tasks in the workflow 
\end{algorithm}

\subsection{Renting VMs by Reserved Cost}
\begin{algorithm}[tb!]

\label{alg:reserved}
\caption{Renting VMs by Reserved Cost using Predicted Data}
\KwIn{\footnotesize $free\_VMs \leftarrow$ list of VMs rented and free, $v^k_i \gets$ $i^{th}$ task of $k^{th}$ workflow, $\mathcal{C} \leftarrow$ total renting cost}
\KwOut{\footnotesize Newly rented VMs by Reserved Cost}

VM Pricing Manager provides information on available reserved and spot cost predictions and corresponding VM configurations

\If{\textnormal{spot price prediction is not available}}{

$Reserved_{Prob} \gets$ predefined threshold for renting a VM using the reserved pricing model

Reserve a VM with probability $Reserved_{Prob}$

}
\Else{
$A \leftarrow$ Number of spot VM arrivals in the batch interval for each VM type

$U \leftarrow$ Number of VM instances needed in the batch interval for each VM type

\If{$A \leq U$}{
Rent VMs of the corresponding types by paying the reserved cost, ensuring that both computation power and memory requirements are satisfied
 
}

}
$\mathcal{C} \gets$ Update the rental cost

Add the newly rented VMs in $free\_VMs$

Return updated $free\_VMs$
\end{algorithm}

We rent VMs in advance based on the predicted workload, adopting a probabilistic strategy for provisioning through the reserved pricing model. While reserved costs are generally stable and reliable, requiring pre-booking, there are scenarios where spot instances may become available at significantly lower prices. In such cases, renting spot VMs in real-time can lead to substantial cost savings. The detailed pseudocode for this strategy is provided in Algorithm \ref{alg:reserved}.

Our approach considers two cases based on the availability of short-term spot instance predictions. In the absence of spot market predictions, VMs are rented based on a probabilistic policy. Specifically, a VM is rented at each decision point with a predefined probability $Reserved_{Prob}$. This strategy intentionally leaves some time intervals without reserved VMs, creating an opportunity to exploit spot instances if they become available later. However, this decision is made probabilistically due to the lack of precise spot availability information (\texttt{line 2-4}).

In contrast, when short-term spot market predictions are available, the decision-making process becomes deterministic. We evaluate the predicted arrivals of spot VMs and assess their capability to meet the computation and memory requirements of upcoming tasks. Based on this information, VMs are reserved only if the predicted spot supply is insufficient to meet the anticipated demand (\texttt{line 5-9}).

After the necessary VMs are rented, the total rental cost $\mathcal{C}$ is updated, and the newly provisioned VMs are added to the $free\_VMs$ pool for task scheduling (\texttt{line 10-12}).

As resources are rented, the VM pool dynamically expands (up-scaling). Each reservation is made for a fixed duration, typically in hours, after which the VM is automatically removed from the pool (down-scaling). At the junction of two consecutive rental periods, for example, if the SCSP rents 10 VMs in one period and requires only 8 in the next, the SCSP avoids fully deallocating all 10 VMs and re-renting 8 new ones. Instead, it renews the rental for 8 existing VMs and releases the remaining 2. This strategy enhances cache retention and reduces cold start overheads, improving execution efficiency.

\subsection{Real-time Renting of VMs by On-Demand or Spot Cost}

In real-time, when renting additional VMs for hosting tasks is needed, two possibilities arise depending on the availability of spot instances. If no spot instances are available, the only option is to rent VMs at the on-demand price, which, although reliable, is significantly more expensive. In such scenarios, the SCSP is compelled to incur the higher cost to maintain service continuity. The detailed pseudocode for this strategy is provided in Algorithm \ref{alg:spot1}.

However, a more dedicated decision process is required if spot instances are available. Spot instances are offered at a fluctuating minimum spot price, denoted as $SP$. The SCSP can place a bid above $SP$ to acquire the spot instance. The instance remains accessible as long as the current market spot price remains below the bid price. For instance, if the SCSP bids $SP+\delta$, the instance remains available until the spot price exceeds $SP+\delta$, beyond which the instance is revoked. While revocation preserves task progress through checkpointing mechanisms, it introduces two critical challenges: (1) computational resources are wasted due to the interruption of ongoing tasks, and (2) deadlines may be missed due to task rescheduling delays.

\begin{algorithm}[tb!]
\label{alg:spot1}
\caption{Renting VMs by On-Demand or Spot Cost using Real-time Data}
\KwIn{\footnotesize $free\_VMs \leftarrow$ list of VMs rented and free, $v^k_i \gets$ $i^{th}$ task of $k^{th}$ workflow, $\mathcal{C} \leftarrow$ total renting cost}
\KwOut{\footnotesize Newly rented VMs by On-Demand or Spot Cost}

VM Pricing Manager provides information on available spot and on-demand costs and corresponding VM configurations

\If{\textnormal{no spot VM is available}}{
Rent VMs by Demand Cost
}

\Else{


Calculate bid price for renting spot VM using \autoref{bid_price}

Rent the spot VM, paying the calculated bid price

}

$\mathcal{C} \gets$ Update the rental cost

Add the newly rented VMs in $free\_VMs$

Return updated $free\_VMs$
\end{algorithm}
To navigate this trade-off, we propose an intelligent bidding strategy that balances the risk of revocation against unnecessary cost increases. Bidding higher prices reduces the likelihood of revocation but increases rental costs, whereas bidding lower prices saves costs but increases the risk of losing the VM. Our approach dynamically adapts the bidding strategy based on the value of scheduled tasks.

To quantify the value of tasks, we define a reward structure for each task based on its importance:

\begin{equation} weights_i^k = l_i^k \cdot e^{\lambda\cdot depth(v_i^k)} \end{equation}

where $depth(v_i^k)$ denotes the depth of task $v_i^k$ in the workflow dependency graph $W^k$, and $l_i^k$ is the computational load (length) of the task. Thus, tasks with greater computational length and greater depth are assigned higher importance.

The overall reward of a workflow is proportionally distributed among its constituent tasks based on their weights:

\begin{equation} rewards_i^k = r^k \cdot \frac{weights_i^k}{ \sum_j^{n^k} weights_j^k} \end{equation}

where $r^k$ represents the total reward assigned to workflow $W^k$, and $n^k$ represents total number of tasks in workflow $W^k$.

To operationalize this mechanism, we maintain a cumulative reward sum ($cumulative\_score$) for each VM type. Each time a task is scheduled, its reward is added to the corresponding cumulative sum for the VM type.

When placing a bid for a spot VM, the bid price is determined based on the cumulative reward associated with that VM type during the expected rental duration, using the following formula:

\begin{equation} bid = DP - (DP - SP) \cdot e^{-\alpha \cdot cumulative\_score} \label{bid_price} \end{equation}

where $DP$ denotes the on-demand price of the VM type, $SP$ is the current spot price of that particular VM type, and $\alpha$ is a tunable parameter that controls the sensitivity of the bidding strategy. This formulation ensures that the bidding price gradually increases with the importance of scheduled tasks, allowing critical tasks to secure spot instances more reliably while minimizing unnecessary cost escalation for less critical workloads.

\subsection{Time Complexity}
The offline scheduling solution is computed before real-time execution using predicted task arrival data. So we focus our attention to the online component and find out the worst-case time complexities for handling spot instance arrivals and task arrivals.

To identify tasks ready for execution, we perform a Depth-First Search (DFS) over the workflow DAGs, identifying all tasks whose dependencies have been satisfied. The complexity of this step is \(O(|\mathcal{W}_b|\cdot (n^k)^2)\), where \(|\mathcal{W}_b|\) denotes the total number of workflows in batch \(b\), and \(n^k\) represents the number of tasks within workflow \(W^k\). Specifically, a single DFS traversal incurs \(O(n^k + (n^k)^2)\) complexity, as the maximum number of edges in the workflow DAG is \(O((n^k)^2)\). The complexity of finding the relative deadline and the relative compute power are $O(n^k)$ and $O(1)$ respectively for each workflow.

After identifying the ready tasks, selecting an appropriate VM from the existing VM pool to schedule each task requires \(O(M_b)\) time, where \(M_b\) is the number of VMs available in the SCSP's pool at batch \(b\). In cases requiring new VM rentals, the complexity reduces to \(O(1)\), since the number of VM types is constant.

Hence, the overall scheduling complexity per batch is: $O(|\mathcal{W}_b|\cdot ((n^k)^2 + n^k + 1) + M_b)$.

\section{Experiments and Results}

We developed a custom-built simulator in Python to evaluate various scheduling strategies for the SCSP. The simulator handles VM provisioning, manages task queues, and tracks rewards associated with the successful completion of workflow executions. 
\subsection{Real-life Dataset}
In our study, we used the Pegasus Workflow Management System to simulate real-world scientific workflows \cite{10.1016/j.future.2014.10.008, pegasus_examples}, accurately modelling task dependencies, execution times, and resource demands. Workflow submission times were randomly distributed over a 20-hour window to reflect dynamic and unpredictable arrivals. To simulate uncertainty in predicted arrival times, we generated a corresponding dataset using a Gaussian error-based approach, where prediction errors were sampled from a Normal distribution with tunable mean and variance. This setup enabled controlled evaluation of scheduling performance under varying levels of prediction accuracy.

For the economic modelling of virtual machine provisioning, we incorporated real-world pricing data from Amazon Web Services (AWS) \cite{aws_c3_8xlarge}. Reserved and on-demand instance costs were obtained from AWS’s publicly available pricing information, enabling evaluation of deterministic pricing strategies. To capture the volatile nature of spot instances, we analyzed historical spot price data obtained from a Kaggle dataset \cite{kaggle_spot_pricing, 10.1007/s10723-023-09676-9}, which provided insight into the fluctuations and pricing patterns over time. The predicted values of spot prices are derived from historical data. We conducted experiments using the set of hyperparameter values listed in \autoref{tab:vm_config}, with 1000 workflows and medium spot VM density availability unless stated otherwise.

\begin{table}[t]
\centering
\caption{VM Configuration and Pricing. Mem denotes VM memory (GiB), CP is compute power (vCPUs $\times$ GHz), OD is on-demand cost, and Res is reserved cost (in \$/hr).}

\setlength{\tabcolsep}{4pt}
\renewcommand{\arraystretch}{0.95}
\begin{tabular}{lccc}
\toprule
\textbf{Type} & \textbf{Mem (GiB)} & \textbf{CP} & \textbf{Price (\$/hr)}, OD / Res  \\
\midrule
c3.large     & 3.76   & 5.6   & 0.105 / 0.073 \\
c3.2xlarge   & 15.04  & 22.4  & 0.420 / 0.292 \\
i3.large     & 15.24  & 4.6   & 0.156 / 0.107 \\
c3.8xlarge   & 60.16  & 89.6  & 1.680 / 1.168 \\
i3.2xlarge   & 60.96  & 18.4  & 0.624 / 0.428 \\
i3.8xlarge   & 243.84 & 73.6  & 2.496 / 1.714 \\
\bottomrule
\end{tabular}
\label{tab:vm_config}
\end{table}





\subsection{State-of-the-art methods}
For a fair and comprehensive comparison, we implemented and adapted baseline approaches addressing similar scheduling challenges, with their details and modifications outlined below.

\subsubsection{FaasCache} FaasCache, proposed by Fuerst and Sharma \cite{10.1145/3445814.3446757}, schedules tasks using a priority function that accounts for the cold start time and popularity of the container image likely to be evicted, as well as the machine's idle duration. This strategy resembles a mix of LRU and LFU-based eviction policy.

\subsubsection{Cost Efficient Workflows-as-a-Service Broker (CEWB)} CEWB introduced by Taghavi \textit{et al.} \cite{10.1007/s10723-023-09676-9}, performs task scheduling and machine provisioning at regular intervals. Machines are categorized by bid price, with on-demand instances considered the most reliable. Deadlines are assigned and tasks are prioritised based on slack time and revocation penalties. High-priority tasks are scheduled on higher-class machines to minimise the risk of revocation and deadline miss. This approach originally did not account for cold start; however, for a fair comparison with our method, we integrated our cold start handling module into it.

\subsection{Evaluating Cold Start Mitigation and Deadline Management at Scale}

\begin{figure}[tb!] 
    \centering 
    \includegraphics[width=1\linewidth]{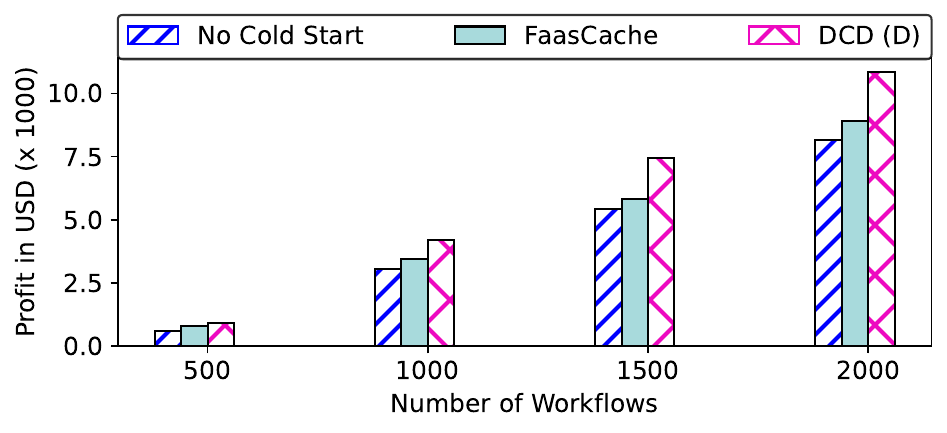} 
    \caption{Impact of Workflow Scaling on Cold Start and Deadline-Aware Scheduling Approaches} 
    \label{fig:Profit_vs_Workflows_Cold_Start} 
\end{figure}

\autoref{fig:Profit_vs_Workflows_Cold_Start} shows the variation in profit across different numbers of workflows for the evaluated cold start mitigation approaches: \textit{No Cold Start}, \textit{FaasCache}, and the proposed \textit{DCD} approach with only demand cost, denoted by \textit{DCD (D)}. In this case, only on-demand cost is considered, as the state-of-the-art approach focuses solely on that pricing model; we maintained this for a fair comparison. \textit{No Cold Start} schedules tasks randomly on available machines. As expected, profit increases with the number of workflows due to the greater opportunity to earn rewards. The proposed approach consistently outperforms both baselines across all workflow counts. This improvement is attributed to its efficient handling of cold starts and effective management of workflow deadlines, which significantly reduces deadline misses. In contrast, \textit{FaasCache} and \textit{No Cold Start} exhibit higher rates of deadline violations, resulting in lower profits. These findings demonstrate the scalability of \textit{DCD} under increasing workload conditions.
\textit{DCD (D)} outperforms \textit{FaaSCache} by up to 20\%.

\subsection{Evaluating Pricing-Based Approaches Across Workflow Scales}
\begin{figure}[tb!] 
    \centering 
    \includegraphics[width=1\linewidth]{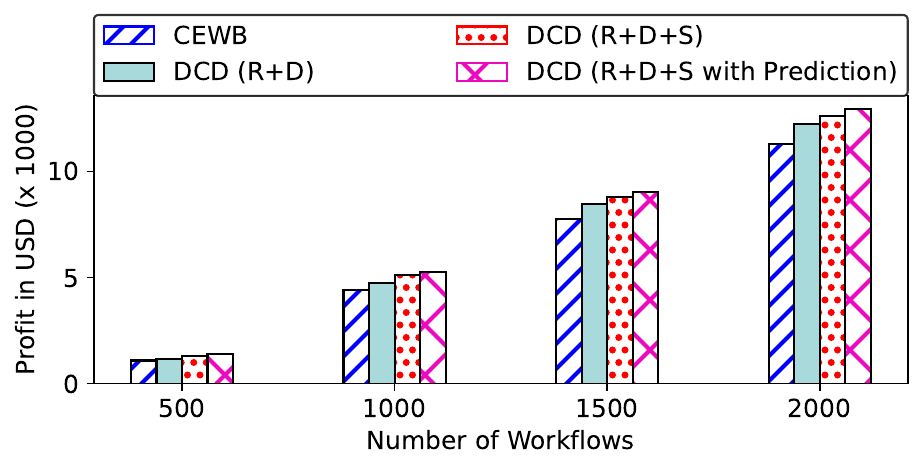} 
    \caption{Impact of Workflow Scaling on the Profitability of Pricing-Based Approaches} 
    \label{fig:Profit_vs_Workflows_Spot} 
\end{figure}

We evaluated the performance of four different approaches: \textit{CEWB}, \textit{DCD (R+D)}, \textit{DCD (R+D+S)}, and \textit{DCD (R+D+S with Prediction)}, across varying numbers of workflows. Here, \textit{DCD (R+D)} refers to the proposed DCD approach utilizing both Reserved VMs ($R$) and On-demand VMs (\textit{D}). \textit{DCD (R+D+S)} extends this by additionally incorporating Spot VMs (\textit{S}) without any spot price prediction. Finally, \textit{DCD (R+D+S with Prediction)} further enhances the model by integrating short-term spot price predictions to improve spot VM utilization decisions. The results, presented in \autoref{fig:Profit_vs_Workflows_Spot}, show that profit increases with the number of workflows, as expected, due to the larger number of opportunities to collect rewards.

All proposed approaches consistently outperform the baseline \textit{CEWB} across all workflow counts. The \textit{DCD (R+D)} approach achieves higher profit by utilizing reserved machines at lower costs. \textit{DCD (R+D+S)} further improves performance by leveraging spot instances, avoiding some reservations, and capitalizing on the lower pricing of spot offerings. The \textit{DCD (R+D+S with Prediction)} variant performs best overall, benefiting from predictive knowledge of spot VMs availability to make more informed decisions on machine provisioning. These results demonstrate the scalability of our approaches under increasing workload conditions. The proposed method, \textit{DCD (R+D+S with Prediction)}, outperforms \textit{CEWB} by up to 15\%.

\subsection{Profit Sensitivity to Spot Instance Density}

\begin{figure}[tb!] 
    \centering 
    \includegraphics[width=1\linewidth]{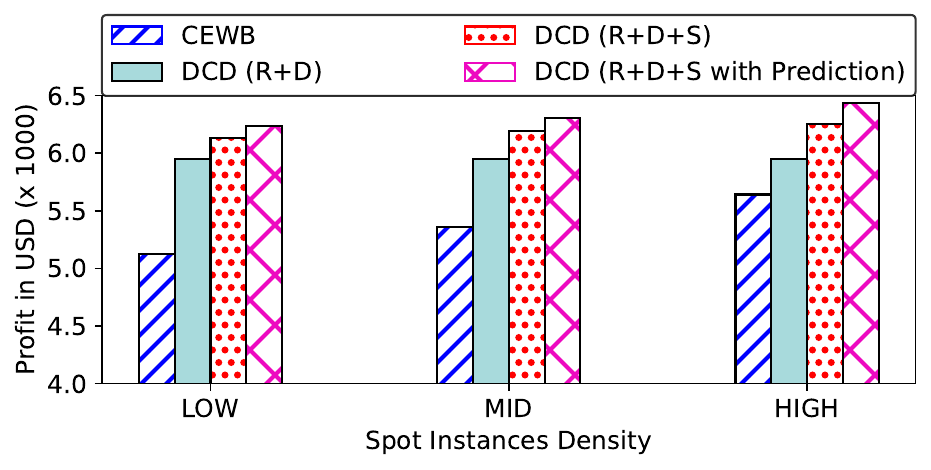} 
    \caption{Impact of Spot Instance Density on the Profitability of Pricing-Based Approaches. \textit{Low} density indicates spot instances are available 10\% of the time, \textit{Mid} indicates 20\% availability, and \textit{High} indicates full (100\%) availability.
} \label{fig:Spot_Density} 
\end{figure}

We evaluate the performance of \textit{CEWB}, \textit{DCD (R+D)}, \textit{DCD (R+D+S)}, and \textit{DCD (R+D+S with Prediction)} under varying availability of spot instances, as shown in \autoref{fig:Spot_Density}. As expected, the profit of \textit{DCD (R+D)} remains constant across different spot instance densities, since it does not utilize spot offerings. In contrast, the profits of \textit{CEWB}, \textit{DCD (R+D+S)}, and \textit{DCD (R+D+S with Prediction)} increase with higher spot instance density, due to their ability to more effectively exploit lower-cost spot resources and reduce overall costs.

However, at low spot instance density, the profit of \textit{CEWB} declines significantly, while our proposed approaches maintain relatively stable performance. This is attributed to \textit{CEWB}'s strong dependence on the availability of spot resources, whereas our approaches are designed to adapt across a range of spot instance densities. These results highlight the robustness of the proposed methods in environments with fluctuating spot availability.

\subsection{Sensitivity to Demand and Reserve Cost Variations}

\begin{figure}[tb!] 
    \centering 
    \includegraphics[width=1\linewidth]{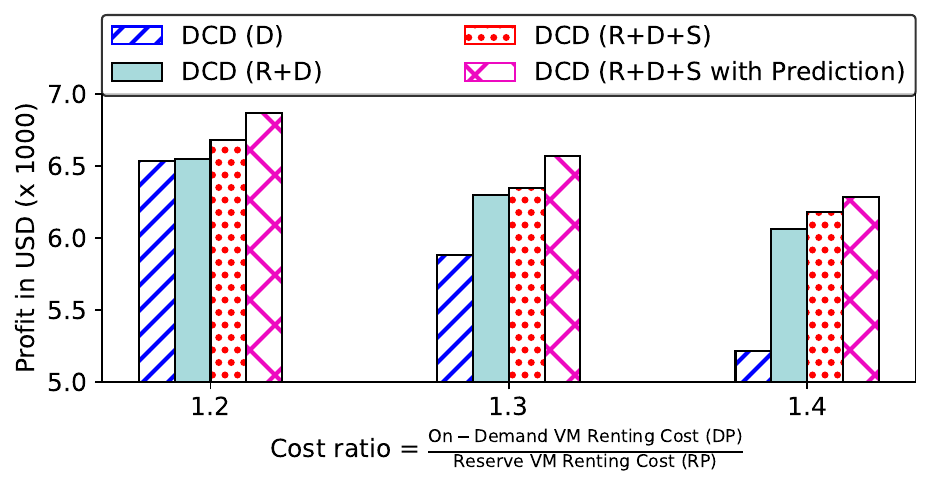} 
    \caption{Profit Variation Across Approaches Under Different Demand-to-Reserve Cost Ratios.} 
    \label{fig:DemandReserveCostVariation} 
\end{figure}

We analyzed the performance of \textit{DCD (D)} (denotes \textit{DCD} approach with only on-demand VM renting), \textit{DCD (R+D)}, \textit{DCD (R+D+S)}, and \textit{DCD (R+D+S with Prediction)} under varying on-demand cost to reserve cost ratios (\textit{i.e} $\frac{DP}{RP}$), while keeping the reserve cost fixed, as shown in \autoref{fig:DemandReserveCostVariation}. We kept the reserved cost (\textit{RP}) fixed and varied the on-demand cost (\textit{DP}). As expected, the overall profit decreases as the on-demand renting cost increases, since the total operational cost rises accordingly.

The most significant decline in profit is observed for \textit{DCD (D)} as it relies exclusively on on-demand instances, and profits go down as on-demand VM rent cost rises. \textit{DCD (R+D+S)} and \textit{DCD (R+D+S with Prediction)} are less sensitive to increases in demand cost compared to \textit{DCD (R+D)}, due to their reduced reliance on on-demand VM renting instances and greater dependence on a mix of spot and on-demand VM renting instances. Due to this sensitivity, both approaches outperform \textit{DCD (R+D)}, demonstrating the cost-effectiveness even under elevated demand pricing scenarios.

\subsection{Impact of Arrival Time Prediction Errors on Scheduling Performance}

\begin{figure}[tb!] 
    \centering 
    \includegraphics[width=1\linewidth]{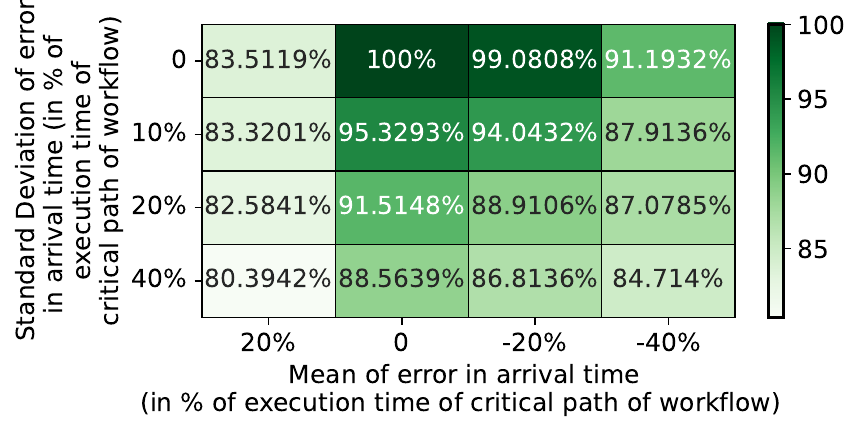} 
    \caption{Sensitivity of Profit to Prediction Error Characteristics in Workflow Arrival Times.} 
    \label{fig:VaryMeanVariance} 
\end{figure}

We analyze the performance of \textit{DCD (R+D+S with Prediction)} under varying levels of prediction error in workflow arrival times, characterized by changes in the mean and standard deviation of the error distribution. \autoref{fig:VaryMeanVariance} presents the resulting profit as a percentage relative to the ideal baseline scenario with perfect predictions (zero mean and zero standard deviation). The means and standard deviations are expressed as percentages relative to the execution time of the tasks. Specifically, for a workflow with an original arrival time $\tau_1$ and an execution time $t$ of their critical path, a reported mean of 40\% implies that the new mean arrival time is shifted to $\tau_1 + 0.4 \times t$. Similarly, the standard deviation is scaled proportionally based on the execution time. 

As anticipated, an increase in the standard deviation of prediction error leads to a decline in profit, attributable to increased uncertainty in workload estimation. A positive mean error results in the system overestimating the actual arrival times, thereby delaying resource provisioning and triggering more frequent use of costly on-demand instances due to missed reservations. Conversely, increasingly negative mean errors cause the system to provision resources prematurely, incurring unnecessary costs from underutilized machines.

\subsection{Effect of Reserve Probability under Prediction Uncertainty}

\begin{figure}[tb!] 
    \centering 
    \includegraphics[width=1\linewidth]{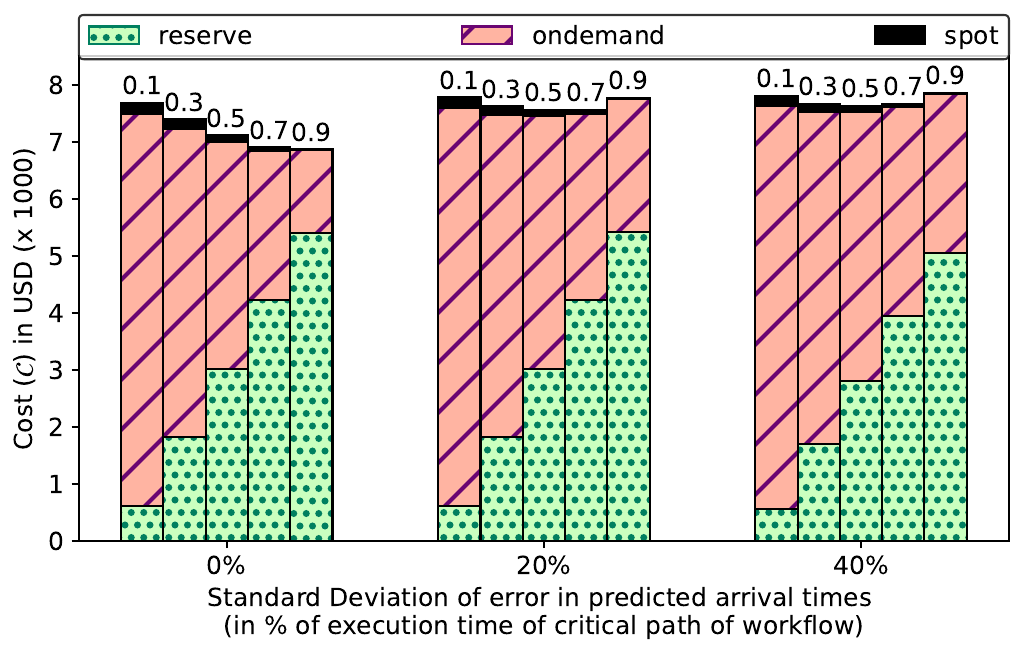} 
     \caption{Impact of Reserved Probability ($Reserved_{Prob}$) under Different Prediction Errors. Lower values on the y-axis indicate better performance in terms of cost (note: unlike other cases where the y-axis represents profit).} 
    \label{fig:ErrorSensitivityAnalysisArrivalRate} 
\end{figure}

We evaluate the performance of \textit{DCD (R+D+S)} in identifying the favorable reserve probability $Reserved_{Prob}$ for VMs renting under varying levels of prediction uncertainty (\textit{i.e.} no spot VM prediction is available beforehand). \autoref{fig:ErrorSensitivityAnalysisArrivalRate} shows the renting cost ($\mathcal{C}$) incurred by \textit{DCD (R+D+S)} across different reserve probabilities ($Reserved_{Prob}$) with varying standard deviations in arrival time prediction error, while keeping the mean error fixed at zero. Like in the previous case, standard deviation is represented in terms of percentage of their execution time of workflow.

As expected, when the standard deviation is zero, indicating perfect prediction accuracy, the cost consistently decreases with increasing reserve probability. This is because accurate predictions allow efficient utilization of reserved instances, reducing reliance on expensive on-demand machines. However, as prediction uncertainty increases (i.e., with higher standard deviation), the favourable reserve probability shifts to a mid-level value. In such cases, reserving all machines becomes inefficient, as it fails to accommodate variability in arrival times. Instead, a balanced strategy that combines partial reservation with opportunistic use of spot instances yields lower overall cost. These results demonstrate the importance of adapting reserve probability ($Reserved_{Prob}$) based on the level of prediction uncertainty.

\section{Conclusion and Future work}
This paper addresses critical challenges in scientific workflow scheduling within cloud environments, emphasizing cold start mitigation and cost-effective management of diverse VM pricing models (reserved, on-demand, and spot instances). We introduced a unified hybrid scheduling framework combining historical workload predictions with real-time adjustments to enhance the profitability of Scientific Cloud Service Providers (SCSPs). Our hybrid two-phase strategy plans renting VM by reserved cost based on workload predictions and spot instance, dynamically adapting to actual demand by provisioning additional on-demand or spot instances, is a new concept for workload scheduling. The framework incorporates cost-aware VM management, featuring a priority-based VM selection and reward-aware spot bidding strategy, effectively balancing cost savings with revocation risk. Extensive experiments with real-world workflows and pricing data confirmed the framework's scalability and robustness, consistently outperforming baseline methods despite substantial prediction errors. Future work includes exploring dynamic hyperparameter tuning and integrating advanced reward mechanisms to optimize profit and efficiency further.

\footnotesize
\bibliographystyle{IEEEtran}
\bibliography{btp}
\end{document}